\documentclass{aa}
\usepackage{graphicx}

\begin{document}
   \title{Twofold effect of Alfv\'en waves on the transverse 
          gravitational instability}

   \author{Adriana Gazol\inst{1} and Thierry Passot\inst{2} }

   \offprints{A. Gazol}

   \institute{Instituto de Astronom{\'\i}a, Universidad Nacional Aut\'onoma 
              de M\'exico, Campus Morelia, \\
              A. P. 72-3 (Xangari), 58089, Morelia, Michoac\'an, M\'exico\\
              \email{a.gazol@astrosmo.unam.mx}
              \and 
              CNRS, Observatoire de la C\^ote d'Azur, B.P.\ 4229, 06304, Nice
              Cedex 4, France\\
              \email{passot@obs-nice.fr}}


   \abstract{
This paper is devoted to the study of the gravitational instability
of a medium permeated by a uniform magnetic field along which a
circularly polarized Alfv\'en wave propagates. We concentrate on the
case of perturbations  purely transverse to the ambient field by means
of direct numerical simulations of the MHD equations and of a linear
stability analysis performed on a moderate amplitude asymptotic model.
The  Alfv\'en wave  provides an extra stabilizing pressure
when the scale of perturbations is sufficiently large or small compared with
the Jeans length $L_J$. However, there is a band of scales around 
$L_J$ for which the Alfv\'en wave is found to have a destabilizing effect.
In particular, when the medium is stable in absence of waves, 
the gravitational instability can develop when the wave amplitude
lies in an appropriate range.
This effect appears to be a consequence of the coupling between
Alfv\'en and magnetosonic waves. The prediction based on a WKB
approach that the Alfv\'en wave pressure tensor is isotropic and
thus opposes gravity in all directions is only recovered for
large amplitude waves for which the coupling between the different
MHD modes is negligible.

}
\titlerunning{Alfv\'en waves effect  on the transverse 
          gravitational instability}
\authorrunning{Gazol \& Passot}
\maketitle

\section{Introduction}\label{intro}
 Formation of the largest clouds in the interstellar medium (ISM) 
certainly rely on ram pressure of large-scale turbulent flows (\cite{paperI}),
whereas molecular clouds most likely owe their stability to internal 
turbulent motions. 
It is thus of importance for the understanding of the ISM dynamics,
to study the gravitational instability in media stirred by MHD turbulence.
 Since the first stability analysis by Jeans
(\cite{Jeans}), there has been a series of studies of the onset of
gravitational instabilities in rotating, magnetized or turbulent
media (e.g.  \cite{chandra61}; \cite{toomre64},
\cite{elme91}). Neutral turbulence was shown to oppose gravity
by a turbulent pressure if its characteristic scale is small enough
compared to the instability wavelength (\cite{chandra51}). 
In comparison with the mechanism that generates density fluctuations,
turbulent pressure is however a higher order effect (\cite{sasao73}).
 Later studies predict a reversal of the Jeans' criterium if the 
turbulent spectrum is shallow enough (\cite{Bona87}; \cite{VazGa}). 
Numerical simulations in two dimensions 
confirm that highly compressible turbulence can have both a
stabilizing effect for long-wavelength perturbations and a
destabilizing one at small scales (\cite{LPP90}).

Consideration of MHD turbulence is crucial in the study of star
formation. As is well known, magnetic fields are ubiquitous in the ISM
and they are believed to be one of the major agents controlling 
star formation (e.g. \cite{Shu87}). If the value of the 
ambient magnetic field (supposed to be relatively smooth over the
large scales) is large enough (sub-critical regime), overall collapse
is hindered and the interstellar matter gathers under a
quasi-static contraction due to ambipolar drift until a super-critical
core is formed : this is the low-mass star formation process. 
On the contrary, if the magnetic field is weak, matter can easily
become gravitationally unstable and high-mass stars form in a rapid 
collapsing process (\cite{mousch}).
In this picture turbulence is considered as an additional
pressure against gravity. This pressure could be provided
by turbulent magnetic field fluctuations.  

In the previous picture the interactions between the turbulent 
motions and the magnetic field are however not consistently taken 
into account  and it is legitimate to
wonder whether this star formation scenario still holds in a fully
turbulent case. The key point in this bi-modal star
formation theory is that magnetic fields always oppose gravitational
instabilities, at least in the linear regime. A uniform magnetic field
is known to increase the critical wavelength for instability by a
factor $({\beta/1+\beta})^{-1/2}$ (where $\beta$ is up to a constant,
the ratio of thermal over
magnetic pressure) if the perturbation is exactly perpendicular to the
ambient field, whereas it has no effect in every other direction.
Support along the magnetic field is provided by Alfv\'en waves (AWs). 
Recent linear stability analyses indeed  show that 
AWs provide an extra pressure for perturbations
parallel to the background field and as a result the Jeans length can be
arbitrarily increased for sufficiently large wave amplitude 
(\cite{Lou1996}; \cite{FH99}). 
Numerical simulations in a slab geometry 
have confirmed this result for a spectrum of AWs 
(\cite{GO96}). 
However, higher dimensional models including decaying MHD turbulence
(e.g. \cite{SOG98}; \cite{OGS99}) have 
shown that in magnetically supercritical
clouds the time for gravitational collapse does not depend on the initial  
level of turbulence. In high resolution three dimensional 
MHD simulations, the presence of short wavelength MHD waves appears to
be able to delay the collapse (\cite{HMLK01}) 

Other studies by \cite{pudritz90} and \cite{Zweibel} use a WKB theory
\cite{Dewar} to argue that the AW pressure tensor is isotropic and
that these waves provide an additional support in every
direction.
The phenomenological argument is based on the observation that the
pressure tensor of the waves $P_W=\frac{\delta B^2}{8\pi}{\bf I}
-\frac{\delta {\bf B}\delta {\bf B}}{4\pi}+\rho \delta {\bf v}\delta
{\bf v}$ reduces to $P_W=\frac{\delta B^2}{8\pi}{\bf I}$ since for a
traveling AW, the equations of motion imply that $\delta {\bf v}=\pm 
\frac{\delta {\bf B}}{(4\pi\rho)^{1/2}}$.
This conclusion is however based on the assumption that the only waves
propagating in the medium are pure AWs. Even though AWs are certainly
the only kind of MHD waves that can propagate over large 
distances due to
their quasi-non-dissipative properties (they are transverse waves),
as soon as they are perturbed, they generate long wavelength
magnetosonic waves. The latter are intrinsically coupled to AWs and
can drastically change the global response of AWs under an external
compression. 

In this paper, we address the simple question of the linear stability
of a self-gravitating medium permeated by a uniform magnetic field
${\cal B}_0$ along which a finite amplitude circularly polarized
AW propagates. In particular, we study the case in which perturbations
are transverse to the uniform magnetic field.
We first present a numerical test showing that for transverse
perturbations, the presence of a longitudinal AW does not always
provide an additional support against gravitational collapse.
In order to explain this result we have done a linear stability
analysis. Due to the presence of several kinds of instabilities taking place
at different scales (e.g. decay instability (\cite{gold78}; 
\cite{derby})) and the coupling
between them, the three dimensional analysis performed on the 
complete MHD equations appears to be cumbersome and physically unclear. 
In fact this problem, which leads to linear equations with
periodically varying coefficients, requires a Floquet analysis.
When the perturbations are parallel to the AW, the infinite hierarchy of
dispersion relations obtained in this way decouples into a set
of physically equivalent dispersion relations (\cite{JH93}). 
However, when perturbations have an arbitrary direction with respect to the
propagation direction of the AW, the modes remain coupled and
the hierarchy of dispersion relations needs to be truncated.
Even in this case the resulting dispersion
relation is very complicated (\cite{VG91} obtain a
78 order polynomial). For this reason we perform a linear stability 
analysis on a reduced description of the MHD equations based on long 
wavelength, small amplitude perturbative expansion (\cite{GPS99}).

The plan of the paper is as follows. In Section 2 we present the numerical
results.  The reduced system of equations used for the theoretical 
interpretation of these results is described in Section 3.
Its derivation is presented in the Appendix, and its 
linear stability analysis in Section 4. Section
5 is a conclusion. 

\section{Numerical Simulations}\label{sec:num}

In this section we present direct numerical simulations of the self-gravitating
2.5D MHD equations
\begin{eqnarray}
&&\partial _{t}\rho + {\bf \nabla}\cdot( \rho{\bf u} )  =
0\label{eq:mhd1s}\\ 
&&\rho ( \partial _{t}{\bf u} + {\bf u}\cdot{\bf \nabla}{\bf u} )  =    
- \frac{\beta}{\gamma}{\bf \nabla}\rho ^{\gamma} + ( {\bf
  \nabla}\times {\bf b} ) \times {\bf b}\nonumber\\
&&\;\;\;\;\;\;\;\;\;\;\;\;\;\;\;\;\;\;\;\;\;\;\;\;\;\;
-\beta J^2\rho\nabla\Phi\label{eq:mhd2s}  \\
&&\partial _{t}{\bf b} - {\bf \nabla}\times ( {\bf u}\times {\bf b}) 
 = 0\label{eq:mhd3s}  \\
&&{\bf \nabla}\cdot{\bf b}  =   0\label{eq:mhd4s}\\
&&\nabla ^2\Phi =\rho -1,\label{poisons} 
\end{eqnarray}
\noindent
where all variables have been adimensionalized using
as velocity  unit, the Alfv\'en speed $v_A={\cal B}_0/(4\pi\rho_0)^{1/2}$,
with ${\cal B}_0$ and $\rho_0$ being  the ambient
magnetic field, chosen to point along the x-axis,
and the average density  respectively.
In equation (\ref{eq:mhd2s}) $\gamma$ denotes the polytropic 
gas constant, $\beta=c_s^2/v_A^2$ with $c_s$ the sound speed, and $J$ 
denotes the Jeans number, ratio of the characteristic scale $L_0$ 
to the Jeans length $L_J=(\pi c_s^2/G\rho_0)^{1/2}$. The simulations have
been performed using a pseudo-spectral code with a resolution 
of $128^2$ grid points, except when otherwise specified.

Previous numerical work in more than one space dimension have mainly
addressed the influence of AWs on the gravitational collapse in a turbulent
context. In order to clearly identify the AW pressure contribution
we choose here to study the gravitational instability close to threshold
in presence of a parallel propagating, right-hand circularly polarized AW which
is an exact solution even in the presence of gravity.
We focus on the case of purely transverse perturbations and thus choose
the longitudinal size of the computational domain $L_x$ to be slightly
smaller than $L_J$ so that the system is stable in the direction along
the uniform magnetic field. The critical transverse size of the domain
is 
$$L_{\perp J}=\frac{2\pi}{J}\left(\frac{1+\beta}{\beta}\right)^{1/2}.$$
In all subsequent simulations we fix
$\gamma=5/3$, $J=1$ and just vary the transverse scale $L_\perp$. The
AW of amplitude $B_0$ reads
$b_y+ib_z=(\omega/k)u_y+iu_z=B_0e^{i(kx-\omega t)}$, $\rho=1$ and $u_x=0$,
where for the nondispersive case $\omega=k$.
The wave number is usually taken as $k=1$ and a noise of 
$10^{-7}$ in the first five Fourier modes is superimposed on the uniform 
density.  

We first consider the case $\beta=0.5$, for which perturbations at
scales $L_\perp>L_{\perp J}=2\pi\sqrt 3$ are gravitationally unstable
in the absence of waves. Choosing $L_\perp/2\pi=1.74$ and $L_x/2\pi=0.948$
a series of six simulations varying the wave amplitude has been performed.
When $B_0=0$ the growth rate of the gravitational instability is
verified to agree  
within 1\%  with the theoretical value $0.068$. Increasing $B_0$ by
steps of 0.1, 
the growth rate is first observed to increase up to $0.107$ for
$B_0=0.3$. It then 
decreases until $B_0=0.5$ for which the system becomes stable. 
This result confirms that a large enough amplitude AW successfully 
stabilizes the medium against the gravitational instability. However
it is surprising that the growth rate increases for small wave
amplitudes.

\begin{figure}
\resizebox{\hsize}{!}
{\includegraphics{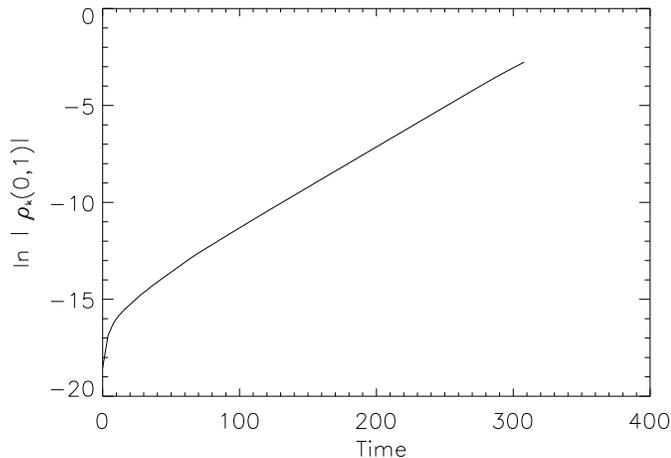}}
\caption[]{$|\ln\rho_k(0,1)|$ as a function of time for 
$\beta=0.5$, $L_x/2\pi=0.95$, $L_\perp/2\pi=1.7$ and $B_0=0.3$}
\label{fig:taux_05}
\end{figure}

\begin{figure}
\resizebox{\hsize}{!}
{\includegraphics{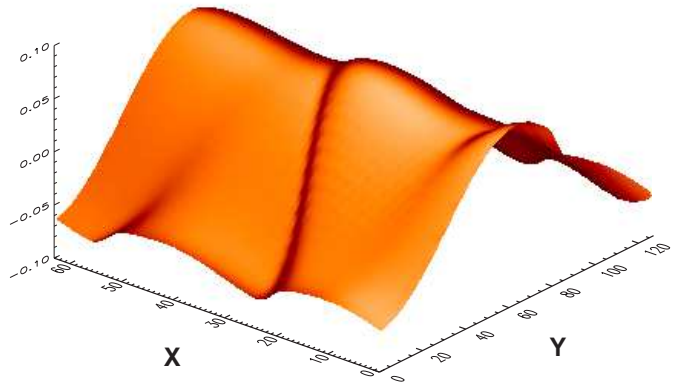}}
\caption[]{Density field fluctuations for $\beta=0.5$, $L_x/2\pi=0.95$, 
$L_\perp/2\pi=1.7$  and $B_0=0.3$, at $t=288$}
\label{fig:den_05}
\end{figure}

\begin{figure}
\resizebox{\hsize}{!}
{\includegraphics{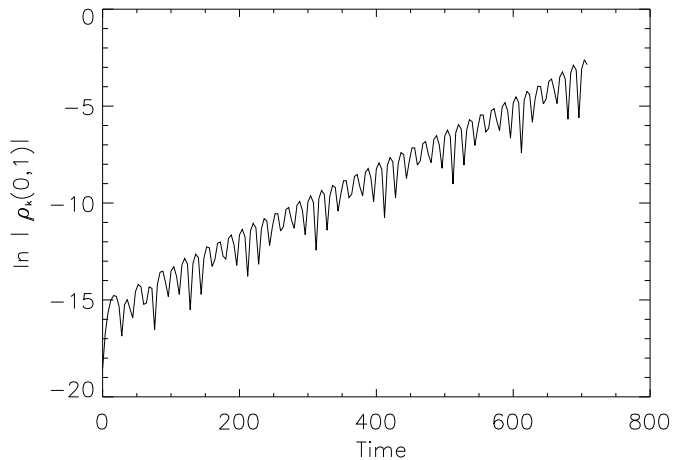}}
\caption[]{$|\ln\rho_k(0,1)|$ as a function of time for 
$\beta=2.0$, $L_x/2\pi=0.7$, $L_\perp/2\pi=1.21$ and $B_0=0.3$}
\label{fig:taux_2}
\end{figure}

This suggests to study the behavior of the gravitationally stable 
system in presence of an AW.
Numerical simulations are now performed with a resolution of
$64\times128$ grid points for $L_x/2\pi=0.95$ and $L_\perp/2\pi=1.7$ 
so that the medium is gravitationally stable.
The system remains stable in presence of the AW as long as the wave
amplitude does not exceeds 0.16 but becomes unstable in the range
of amplitudes $[0.18,0.35]$. Figure \ref{fig:taux_05} shows the typical 
temporal behavior of $|\ln\rho_k(0,1)|$, where $\rho_k(k_x,k_y)$ denotes the
Fourier coefficient of the density, for an unstable simulation.
In Fig. \ref{fig:den_05}  the density field fluctuations $\rho-1$  
resulting from  the same simulation are displayed at $t=288$.
In addition to the concentration of mass in the transverse direction,
a weak longitudinal modulation is also visible.
Another set of simulations with $\beta=2$, $L_x/2\pi=0.7$ and 
$L_\perp/2\pi=1.21$, shows the same behavior with respect to
the gravitational instability but in this case, 
the growth of the gravitational instability, shown in Fig. \ref{fig:taux_2},
is oscillatory.

Similar results are also obtained when choosing the AW wave number on the 
third Fourier mode. However in the case $\beta<1$, larger wave numbers
or smaller values of $\beta$ lead to stronger  longitudinal modulation 
resulting from the coexistence with the decay instability.

In the next two sections we present an analytical interpretation of the
previous results.  
  
\section{Reduced description}\label{sec:reddesc}

 As mentioned in Sect. \ref{intro} the analytical study of the 
transverse gravitational stability of the system is a very complex problem.
Its solution would require numerical simulations of the same complexity
as the ones done in the previous section.
For this reason and in order to gain some physical insight, we use a reduced 
description adapted to study the nonlinear
dynamics of an AW propagating along a strong uniform field coupled with a
small amplitude MHD flow in planes perpendicular to this field.
This reduced description has been derived in the absence of self-gravity by
Gazol et al. (1999) (see also \cite{Champ4}) for two different regimes 
depending on the
values of $\beta$ and its derivation is presented in Appendix \ref{appI} for  
the self-gravitating case. A verification of this asymptotic description
has been performed in the context of Hall MHD for the development of transverse
instabilities on a moderate amplitude AW (\cite{LPS02b}). 
Here we only consider  the case $\beta\neq 1$, for
which the reduced equations have been extended to  
take  into account the coupling with magnetosonic waves.
The case where $\beta$ is close to the resonance $\beta=1$ is more delicate
as both the AW and the sonic wave must be considered on an equal footing
and obey nonlinear equations.
The derivation is based on the existence of a small parameter $\epsilon$,
squared ratio of the wave amplitude to the value of the strong 
uniform magnetic field, that identifies with the
squared Alfvenic Mach number, $M_A^2=u_0^2/v_A^2$, where $u_0$ is a 
typical velocity fluctuation.
In the frame of reference of the AW, 
a stretching of coordinates allows to study non-linear effects
on long times, $\tau=\epsilon^2 t$  and large scales, $(\xi,\eta,\zeta)=
(\epsilon(x-t),\epsilon^{3/2}y, \epsilon^{3/2}z)$, for both the AW and
the transverse MHD flow. The scaling and the resulting
equations are given in Appendix \ref{appI}. The coupling between the AW and 
the transverse hydrodynamics is accomplished through the introduction of
mean transverse velocity and magnetic fields which describe the slow 
transverse dynamics and  result from the average over the direction of the 
uniform magnetic field.  

Within the framework of this description, where the small-scale
magnetosonic waves have been filtered, it is possible to
study the interplay between the AW and the
large-scale transverse gravito-magneto-acoustic mode 
in isolation from the other small-scale instabilities.
The scaling is chosen in order to ensure that the transverse scales
are of the same order as the Jeans length.

The reduced description presented in Appendix \ref{appI} includes the Hall
effect, which causes the AW to become dispersive. This effect is probably 
unimportant in must situations of interest although it can become relevant 
in presence of dust (\cite{ruda01}).

\section{Linear stability analysis}
A  plane wave $B_0e^{(i(k\xi-\omega\tau))}$ propagating parallel to the 
uniform magnetic field in a homogeneous medium, is an exact solution to
equations (\ref{sf:3})-(\ref{proc:45}) when its frequency and 
wave number are related by $\omega+k^2/2R_i=0$. The dispersive effect is
kept for generality but it does not affect the analysis. 
The wave is perturbed  as
\begin{equation}
\widetilde b=B_0(1+A)\exp i(k\xi-\omega\tau +\phi),
\label{pert-1}
\end{equation}
where the amplitude, $A=A(\eta,\zeta, \tau)$, and
the phase, $\phi=\phi(\eta,\zeta, \tau)$ modulations are
real and taken independent of the longitudinal variable. 
Such a perturbation assumes that the wave remains circularly
polarized, a correct assumption in the context of a linear stability
analysis, but that ceases to be valid when strong modulations are 
considered (\cite{Champ3}).
  
It is easy to see from equation (\ref{sf:8}) that if the mean
transverse magnetic field $\bar b$ does not exist initially, as we assume,
it is not generated by linear perturbations.
The other fields are written as $p=p^{(0)}+p^{(1)}$,
where the perturbation $p^{(1)}$ separates into oscillating $\widetilde p^{(1)}$
and mean $\bar p^{(1)}$
contributions taken in the form   
$$
\widetilde p^{(1)}=\widetilde p e^{i(k\xi -\omega\tau+\phi)}+c.c,\label{pero}
$$
and
$$
\bar p^{(1)}= \bar p(\eta, \zeta, \tau ),\label{perm}
$$
respectively.
Note that the amplitude of fluctuating perturbations, $\widetilde p$,
 is a complex number and varies only in the transverse directions,
$\widetilde p=\widetilde p^R(\eta, \zeta,\tau)+i\widetilde p^I(\eta, \zeta,\tau)$ while $\bar p^{(1)}$ is a real quantity.
After linearizing and projecting on the first longitudinal Fourier mode we
 get the
 following linear system 
\begin{eqnarray}
&&B_0(i\partial_\tau \phi +\partial_\tau  A)
+ikB_0\left(\bar u_{x}+\frac{\bar b_{x}}{2}-\frac{\bar \delta}{2}\right)\nonumber\\
&&\quad\quad -\frac{1}{2}\partial_\perp(\beta\widetilde\rho+\widetilde b_{x}+\beta j^2\widetilde\Phi)=0\label{el1:1}\\
&&\partial_\tau \bar u+\partial_\perp\left(B_0^2A+\beta\bar\delta+(\beta+1)\bar b_{x}+\beta j^2\bar\Phi\right)=0\label{el1:2}\\
&&\widetilde u_{x}-\widetilde\delta=0 \label{el1:3}\\
&&\widetilde u_{x}-\beta\widetilde\delta-\beta\widetilde b_{x}
-\beta j^2\widetilde\Phi=0 \label{el1:5}\\
&&ik\widetilde b_{x}+\frac{B_0}{2}(\partial_\perp^*A+i\partial_\perp^*\phi)
=0\label{el1:7}\\
&&\partial_\tau \bar\delta
+\frac{B_0}{2}\left(\partial_\perp(\widetilde u_{x}-\widetilde\delta)
+\partial_\perp^*(\widetilde u_{x}^*-\widetilde \delta^*)\right)=0\label{el1:4}\\
&&\partial_\tau \bar u_{x}
-\frac{B_0}{2}\left(\partial_\perp(\widetilde b_{x}+\widetilde u_{x})+
\partial_\perp^*(\widetilde b_{x}^*+\widetilde u_{x}^*)\right)
=0
\label{el1:6}\\
&&\partial_\tau \bar b_x
+\frac{1}{2\epsilon}(\partial_\perp\bar u^*+\partial_\perp^*\bar u)\nonumber\\
&&\quad\quad -\frac{B_0}{2}\left(\partial_\perp(\widetilde b_{x}+\widetilde u_{x})+
\partial_\perp^*(\widetilde b_{x}^*+\widetilde u_{x}^*)\right)
=0\label{el1:8}\\
&&-\frac{k^2}{\epsilon}\widetilde\Phi
+(\partial_{\eta\eta}+\partial_{\zeta\zeta})\widetilde\Phi
=\widetilde\delta+\widetilde b_{x}\label{el1:9}\\
&&(\partial_{\eta\eta}+\partial_{\zeta\zeta})\bar\Phi
=\bar\delta+\bar b_{x}.\label{el2:10}
\end{eqnarray}
Note that a truncation in Fourier space is also necessary in this case. 
The order $\epsilon$ contribution in the
equation for the mean longitudinal magnetic field $\bar b_x$, is kept in order
to allow transverse magnetosonic waves to propagate.
From equations (\ref{el1:3}), (\ref{el1:5}), (\ref{el1:4}) and (\ref{el1:9}) we have
$$
{\widetilde\delta}={\widetilde u}_{x}\;\;\;\;\;\; 
{\bar\delta}=0.
$$
By separating real and imaginary parts of the remaining equations
and then assuming that the
oscillating, $\widetilde p^R$  $\widetilde p^I$, and non-oscillating, $\bar p^{(1)}$,
 perturbations have the form
\begin{equation}
p =\hat p e^{i(K_\eta\eta+K_\zeta\zeta-\Omega \tau)},\label{perhat}
\end{equation}
we obtain an algebraic system of equations from which we find
\begin{eqnarray}
&&\hat{\widetilde u}_{x}=C_{\epsilon}\hat{\widetilde b}_{x}\\
&&\hat{\widetilde \Phi}=\frac{1+C_\epsilon}{\frac{k^2}{\epsilon}+K_\perp^2}\hat{\widetilde b}_x
\end{eqnarray}
\begin{eqnarray}
\hat{\widetilde b}_{x}^I&=&\frac{iB_0}{2k}(K_\eta\hat{A}+K_\zeta\hat{\phi})\nonumber\\
\hat{\widetilde b}_{x}^R&=&\frac{iB_0}{2k}(K_\zeta\hat{A}-K_\eta\hat{\phi})\label{bx0}
\end{eqnarray}
\begin{eqnarray}
\hat{\bar u}_{y}&=&\frac{K_\eta}{\Omega}\left(B_0^2\hat{A}
+\left(\beta+1-\frac{\beta j^2}{K_\perp^2}\right)\hat{\bar b}_{x}\right)
\nonumber\\
\hat{\bar u}_{z}&=&\frac{K_\zeta}{\Omega}\left(B_0^2\hat{A}
+\left(\beta+1-\frac{\beta j^2}{K_\perp^2}\right)\hat{\bar b}_{x}\right)
\label{u}
\end{eqnarray}
\begin{equation}
\hat{\bar u}_{x}=
\frac{iB_0^2 C_{\epsilon 1}}{2k\Omega}K_\perp^2\hat{\phi}
\label{barux}
\end{equation}  
\begin{equation}
\hat{\bar \Phi}=-\frac{\hat{\bar b}_{x1}}{K_\perp^2},
\label{barPhi}
\end{equation} 
where $K_\perp^2=K_\eta^2+K_\zeta^2$, 
\begin{equation}
C_\epsilon=\frac{\beta-\left(\frac{\beta j^2}{k^2/\epsilon+K_\perp^2}\right)}
       {1-\beta+\left(\frac{\beta j^2}{k^2/\epsilon+K_\perp^2}\right)},
\end{equation}
and
\begin{equation}
C_{\epsilon 1}=1+C_{\epsilon}=\left(1-\beta+\frac{\beta j^2}{k^2/\epsilon+K_\perp^2}\right)^{-1}.
\end{equation} 
Equations (\ref{el1:1}) and (\ref{el1:8}) lead to
\begin{eqnarray}
&&i\Omega \hat A+\frac{C_{\epsilon 2}}{4k}K_\perp^2\hat \phi=0\\
&&-i\Omega\hat\phi+i\frac{C_{\epsilon 1}}{2\Omega}B_0^2K_\perp^2\hat\phi
+\frac{k}{2}\hat{\bar b}_x
   +\frac{C_{\epsilon 2}}{4k}K_\perp^2\hat A=0\\
&&-i\Omega\hat{\bar b}_x+i\frac{K_\perp^2}{\epsilon\Omega}\left(B_0^2\hat A
+\left(1+\beta-\frac{\beta j^2}{K_\perp^2}\right)\hat{\bar b}_x\right)
\nonumber\\
&&\quad\quad-\frac{C_{\epsilon 1}}{2k}B_0^2K_\perp^2\hat\phi=0,
\end{eqnarray}
with $C_{\epsilon 2}=\beta C_{\epsilon 1}(1-j^2/(k^2/\epsilon+K_\perp^2))+1$. 
Rescaling the uniform magnetic field and 
the frequency of the perturbations by a factor $\epsilon^{-1/2}$, and the 
Alfv\'en  wave number $k$ by a factor $\epsilon^{1/2}$, i.e. returning to the
original variables, the coefficients
$C_{\epsilon}$, $C_{\epsilon 1}$, and $C_{\epsilon 2}$ become 
independent of $\epsilon$. We denote the unscaled coefficients by  
$C$, $C_1$, and $C_{2}$, respectively.
The dispersion relation, also independent of $\epsilon$, reads 
\begin{eqnarray}
&&\Omega ^4
\nonumber\\
&&-\Omega^2K_\perp^2\left(
\frac{C_2^2 K_\perp^2}{16k^2}+\frac{3}{4} B_0^2C_1
+(\beta+1-\frac{\beta j^2}{K_\perp^2})\right)
\nonumber\\
&&+\left(\beta+1-\frac{\beta j^2}{K_\perp^2}\right)\nonumber\\
&&\left( \frac{C_2^2K_\perp^6}{16 k^2}
  +B_0^2\frac{C_1 K_\perp^4
(\beta+\frac{3}{4}-\frac{\beta j^2}{K_\perp^2})}{2
(\beta+1-\frac{\beta j^2}{K_\perp^2})} \right)  =0.
\label{rd2}
\end{eqnarray} 
It can be rewritten as
\begin{eqnarray}
\Omega^2&&=
K_\perp^2\left(\beta+1-\frac{\beta j^2}{K_\perp^2}\right)\nonumber\\
&&+ \frac{B_0^2}{2}K_\perp^2 C_1
\left( \frac
{\frac{3}{2}\Omega^2-K_\perp^2(\beta+\frac{3}{4}-\frac{\beta j^2}{K_\perp^2})}
{\Omega^2-\frac{K_\perp^4C_2^2}{16k^2}}
\right ).
\label{rd_int}
\end{eqnarray} 

From equation (\ref{rd2}) it is possible to check that for $C_1>0$,
$\Omega^2$ is always real, so that stability is governed by the sign of the 
last term of this equation. When the system is stable in absence of the AW, 
i.e. 
$K_\perp > K_{\perp J}=j(\beta/1+\beta)^{1/2}$, 
it can be seen that as $K_\perp<K_{\perp c1}$, where
\begin{equation}
K_{\perp c1}^2=\frac{\beta j^2}{\beta+3/4},
\end{equation}
and when $B_0$ exceeds a critical amplitude $B_{0c}$, the value for which 
the last term of equation (\ref{rd2}) vanishes, the system becomes unstable.

 The first term in the right hand side of equation (\ref{rd_int}) contains 
the contribution of the uniform magnetic field, whereas the second one is the
contribution arising from the AW. When the system is unstable in 
absence of the AW, there is another critical wave number $K_{\perp c2}$ 
for which the second term vanishes. It reads 
\begin{equation}
K_{\perp c2}^2=\frac{\beta j^2}{\beta+3/2}.
\end{equation}
Always assuming $C_1>0$, for perturbations with wave numbers $K_\perp$ smaller 
than (respectively larger than) $K_{\perp c2}$, the AW has a stabilizing 
(respectively destabilizing) effect.  
Thus there exists a band of perturbation wave numbers 
$[K_{\perp c 2},K_{\perp c 1}]$ 
around $K_{\perp J}$, for which the presence of the AW has a destabilizing 
effect.
\begin{figure}
\resizebox{\hsize}{!}
{\includegraphics{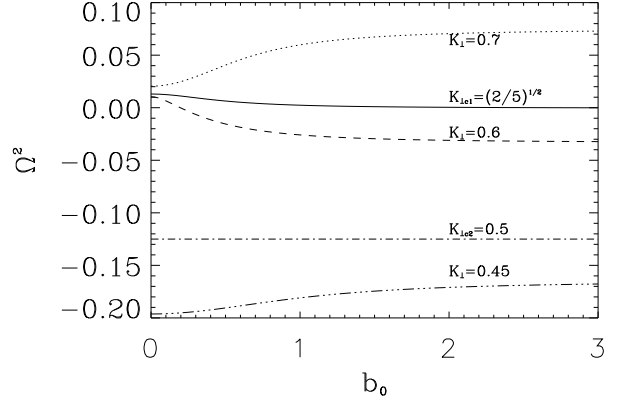}}
\caption[]{Effect of varying the perturbation
wave number near critical values $K_{\perp c1}=(2/5)^{1/2}$ and 
$K_{\perp c2}=0.5$ for  $\beta=0.5$, $j=1$ and $k=1.526$}
\label{fig:omega_05}
\end{figure}
In Fig. \ref{fig:omega_05} we display the root $\Omega^2$ corresponding to the
gravito-magneto-acoustic branch, as a  
function of the wave amplitude $B_0$ for $\beta =0.5$, $j=1$, $k=1.0526$ 
and five different values of $K_\perp$ around $K_{\perp c1}=(2/5)^{1/2}$ and 
$K_{\perp c2}=0.5$ (note that for $\beta<1$, $C_1>0$). It is observed that for 
$K_\perp>K_{\perp c 1}$ (dotted line) the medium remains stable, whereas 
for  $K_\perp<K_{\perp c 1}$ (dashed line) the medium becomes
unstable when the AW amplitude is large enough. 
For $K_\perp<K_{\perp c 2}$ (double-dotted dashed line), 
the medium 
remains unstable but the presence of the AW has a stabilizing effect. 
The destabilizing effect for $K_{\perp c2}<K_{\perp}< K_{\perp c1}$ 
is even more pronounced for larger wave numbers
$k$ although it quickly saturates as $k$ increases.

It is now interesting to discuss whether 
the previous results can be interpreted in terms of an AW pressure.
In the case of quasi-uniform perturbations, i. e. $K_\perp\ll k$
and $K_\perp\ll \Omega$, the dispersion relation (\ref{rd_int}) 
approximates to
\begin{eqnarray}
\Omega^2\approx
K_\perp^2\left(\beta+1-\frac{\beta j^2}{K_\perp^2}\right)
+ \frac{3}{4}B_0^2 C_1 K_\perp^2.
\label{omega_unif}
\end{eqnarray} 
Assuming a polytropic dependence of the magnetic pressure on the density 
in the form $|b|^2/2\propto\rho^\gamma_w$, linearizing equations 
(\ref{eq:mhd1s})-(\ref{eq:mhd2s}), and taking perturbations 
in the form (\ref{perhat}) it is easy to see that for $\beta\rightarrow 0$,
$\gamma_w=3/2$. This is the  value obtained by \cite{Zweibel}
in the same regime. The factor $C_1$ ($C_1=(1-\beta)^{-1}$
as $\epsilon\rightarrow 0$) arises from
the coupling of the AW with long-wavelength magnetosonic waves excited
by the perturbations to the AW.   

When $\Omega$ is real and $\Omega\ll 1$, i.e. close to the stability 
transition for $C_1>0$, the dispersion relation approximates to 
\begin{eqnarray}
&&\Omega^2\approx
K_\perp^2\left(\beta+1-\frac{\beta j^2}{K_\perp^2}\right)\nonumber\\
&&\quad\quad
+8 B_0^2 k^2\frac{C_1}{C_2^2}\left(\beta+\frac{3}{4}
-\frac{\beta j^2}{K_\perp^2}\right).
\end{eqnarray} 
The second term on the right hand side of previous equation is negative
when $K_\perp<K_{\perp c1}$ and overcomes the first term on the right hand side
when $B_0>B_{0c}$. In this case the effect of the AW cannot be modelled by
a pressure.
\begin{figure}
\resizebox{\hsize}{!}
{\includegraphics{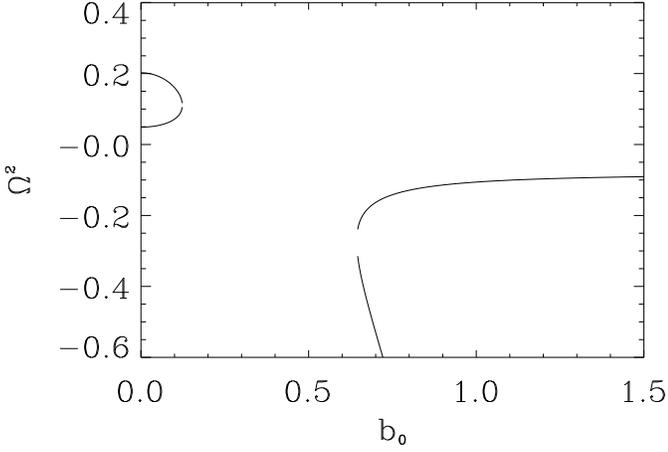}}
\caption[]{Effect of varying
the perturbation wave number for $\beta=2$, $j=1$, $K_\perp=0.826$ 
and $k=1.4286$}
\label{fig:omegap}
\end{figure}

When $C_1<0$, $\Omega^2$ can be either real or complex, depending
on the values of $B_0$, $k$, $K_\perp$ and $j$. 
As an example we choose the  parameters used in the 
second simulation discussed in section \ref{sec:num}, for which
$C_1=-3.73$. In  Fig. \ref{fig:omegap}, $\Omega^2$ is plotted as a function of
$B_0$, when it is real. It can be seen that for $B_0=0.3$, $\Omega^2$ is 
complex, leading to the oscillatory instability observed in Fig.
\ref{fig:taux_2}.   
Note that in addition to the gravitational instability, 
when $\beta>1$, AWs are  unstable with respect to the absolute filamentation 
instability (\cite{Champ1}; \cite{LPS02a}), which
leads to the amplification of the transverse magnetic field in filaments 
aligned with the uniform field.

\section{Conclusion}\label{dis} 

In this paper we have shown that a circularly polarized AW propagating along a
uniform magnetic field modifies the transverse gravitational instability in
two different ways.
When $\beta$ is not too large ($C_1>0$), the AW provides an extra pressure if
the perturbation wave number $K_\perp$ lies outside a band 
$[K_{\perp c 2},K_{\perp c 1}]$. On the other hand, when $K_\perp$ lies
in this band a moderate amplitude AW leads to an increase of the gravitational
instability growth rate or even to the destabilization of a medium
stable in absence of waves.

The asymptotic model used in this paper does not allow to study the
effect of large amplitude waves but it successfully explains direct MHD 
simulations results for moderate amplitude waves.
In addition these simulation show that at large amplitude an AW is able
 to stabilize a
gravitationally unstable medium as predicted by a theory based
on a WKB approach (McKee and Zweibel 1995). The results obtained in
the present paper concerning the destabilizing effect of moderate amplitude 
waves arise from the coupling between the AW and the transversely propagating 
magneto-acoustic mode. This effect, which becomes less relevant
for large amplitude AWs, cannot be recovered by the WKB theory, 
which treats the different MHD modes separately.

The present paper only addresses the linear regime.
The question arises whether the instability triggered by the AW
saturates early or leads to a runaway situation. 
Numerical simulations of the nonlinear regime, which require the presence
of dissipation, are planned.   
 It could also be of interest to perform 
numerical simulations in the regime of weak Alfv\'en wave turbulence,
a strongly turbulent situation being more difficult to analyze.
  
\begin{acknowledgements}
We are thankful to P.L. Sulem for useful discussions. 
This work has received partial financial support from the french national
program PCMI (CNRS), DGAPA IN115400 grant and  CONACYT 36571-E grant.
\end{acknowledgements}

\appendix\section{Reduced equations}\label{appI}

The derivation of the reduced equations is here performed for a 
self-gravitating fluid in the Hall-MHD approximation where equation
(\ref{eq:mhd3s}) is replaced by
\begin{equation}
\partial _{t}{\bf b} - {\bf \nabla}\times ( {\bf u}\times {\bf b}) 
 =  - \frac{1}{ R_i} \mbox{\boldmath $\nabla$}
{\mbox{\boldmath $\times$}} \left( \frac{1}{\rho}  ( \mbox{\boldmath $\nabla$}
{\mbox{\boldmath $\times$}} {\bf b} )  
{\mbox{\boldmath $\times$}} {\bf b} \right). \label{eq:mhd3}  
\end{equation}
\noindent
The last term of this equation, is the Hall term associated with
the generalized Ohm's law and $R_i$ denotes the non dimensional 
ion-gyromagnetic frequency. When the Alfv\'en wavelength is close to
the ion inertial length, this term becomes relevant and makes the
AW dispersive. 

In order to perform the asymptotic expansion we define the stretched variables
$\xi=\epsilon (x-t)$, $\eta=\epsilon^{3/2} y$,  $\zeta=\epsilon^{3/2} z$,
and $\tau = \epsilon^2 t$ and expand
\begin{eqnarray}
&&\rho  =  1 + \epsilon \rho _{1} + \epsilon ^{2}\rho
_{2} + + \epsilon ^{3}\rho _{3}\cdots  \\  
&&u _{x}  =  \epsilon u _{{x}_{1}} + \epsilon ^{2}u
_{{x}_{2}} + \epsilon ^{3}u _{{x}_{3}}+ \cdots  \\ 
&&b _{x}  =  1  + \epsilon b_{{x}_{1}} + \epsilon
^{2}b_{{x}_{2}} +  \epsilon ^{3}b_{{x}_{3}}+\cdots  \\
&&u _{y}   =  \epsilon^{1/2}(  u_{{y}_{1}} +
\epsilon u_{{y}_{2}} + \epsilon ^{2}u_{{y}_{3}} + \cdots) \\
&&u _{z}   =  \epsilon ^{1/2}( u_{{z}_{1}} +
\epsilon u_{{z}_{2}} +\epsilon ^{2}u_{{z}_{3}} + \cdots) \\
&&b _{y}   =\epsilon^{1/2}( b_{{y}_{1}} +
\epsilon b_{{y}_{2}} +\epsilon ^{2}b_{{y}_{3}}  + \cdots) \\
&&b _{z}   =  \epsilon ^{1/2}(b_{{z}_{1}} +
\epsilon b_{{z}_{2}}  +\epsilon ^{2}b_{{z}_{3}}  + \cdots) \\
&&\Phi=\frac{1}{\epsilon^2}( \Phi_0+\epsilon\Phi_1 +\epsilon^2\Phi_2+ \cdots).
\label{dev}
\end{eqnarray}
We also define $J\propto j\epsilon^{3/2}$, ensuring that the typical 
transverse scales are comparable to the Jeans length. When expanding the 
Poisson equation in  powers of $\epsilon$, it can be rewritten as  
\begin{equation}
\left(\frac{1}{\epsilon}\partial_{\xi\xi}+\partial_{\eta\eta}
+\partial_{\zeta\zeta}\right)\Phi=\rho,
\end{equation}
where $\rho=\rho_1+\epsilon\rho_2$ and $\Phi=\Phi_0+\epsilon\Phi_1$.

We then expand equations (\ref{eq:mhd1s}),(\ref{eq:mhd2s}),(\ref{eq:mhd3})
and (\ref{eq:mhd4s}) in powers of 
$\epsilon$. 
At order $\epsilon ^{3/2}$, we have
\begin{equation}
\partial _\xi u_1+\partial _\xi b_1=0,
\label{o3/2:1}
\end{equation}
where the transverse fields are given by $b=b_y+ib_z$ and $u=u_y+iu_z$.
%
%
In order to include a coupling between the Alfv\'en waves and the
transverse hydrodinamic motions, we include a mean contribution, 
corresponding to an average over the
$\xi$ variable and denoted by an over-line, in the transverse components of 
the velocity and magnetic fields
\begin{eqnarray}
b_1=\widetilde b_1(\xi,\eta,\zeta,\tau)+\bar b_1(\eta,\zeta,\tau),\\
u_1=\widetilde u_1(\xi,\eta,\zeta,\tau)+\bar u_1(\eta,\zeta,\tau).
\label{cm}
\end{eqnarray}
Equation (\ref{o3/2:1}) implies that the fluctuating parts (denoted by
tildes) satisfy
\begin{equation}
\widetilde u_1=-\widetilde b_1.
\label{o3/2:2}
\end{equation}
%

At order   $\epsilon ^{2}$, we obtain 
\begin{eqnarray}
&&-\partial _\xi\widetilde\rho _1+\partial _\xi \widetilde u_{x1} 
+\partial _\eta(\widetilde u_{y1}+\bar u_{y1})\nonumber\\
&&\quad\quad +\partial _\zeta(\widetilde u_{z1}+\bar u_{z1})=0
\label{o2:1}\\
&&-\partial _\xi \widetilde b_{x1}+\partial _\eta(\widetilde u_{y1}+\bar u_{y1})
+\partial _\zeta(\widetilde u_{z1}+\bar u_{z1})=0\label{o2:2}\\
&&\partial _\xi\left(-\widetilde u_{x1}+\beta(\widetilde\rho _1+j^2\widetilde\Phi)+\frac{|\widetilde b_1 +
\bar b_1|^2}{2}\right)=0\label{o2:3}\\
&&\partial_\xi\widetilde b_{x1}+\partial_\eta(\widetilde b_{y1}+\bar  b_{y1})
+\partial_\zeta(\widetilde b_{z1}+\bar  b_{z1})=0.\label{o2:4}
\end{eqnarray}
Using equation (\ref{o3/2:2}) and separating the mean values and
fluctuations in equation (\ref{o2:2}), we get
\begin{eqnarray}
&&\partial _\xi \widetilde b_{x1}+\partial _
\eta\widetilde  b_{y1}+\partial _\zeta\widetilde b_{z1}=0
\label{o2:4a}\\
&&\partial_\eta\bar u_{y1}+\partial _\zeta\bar u_{z1}=0.
\label{o2:4b}
\end{eqnarray}
The transverse velocity field is thus incompressible to leading order.
Defining $\partial_\perp = \partial_{\eta} + i \partial_\zeta$, 
Eqs.  (\ref{o2:1}) and (\ref{o2:4}) then rewrite
\begin{eqnarray}
&&\partial _\xi(-\widetilde\rho _1+\widetilde u_{x1}+\widetilde b_{x1})=0
\label{o2:5}\\
&&\partial_\perp^*\bar b_1+\partial_\perp\bar b_1^*=0.\label{o2:6}
\end{eqnarray}
%

At order $\epsilon ^{5/2}$, separating in the equations for $b_1$
the mean and fluctuating parts, we obtain
\begin{equation}
\partial _\tau \bar b_1
-\frac{1}{2}\partial_\perp(\bar u_1\bar b_1^*-\bar u_1^*\bar b_1)=0
\label{o5/2:2a}
\end{equation} 
and
\begin{eqnarray}
&&\partial _\tau  \widetilde b_1-\partial _\xi (\widetilde b_2 +\widetilde u_2)
+(\bar  u_{x1}+\bar b_{x1})\partial _\xi \widetilde b_1\nonumber\\
&&\quad\quad
+\partial _\xi\left(\widetilde b_1(\widetilde u_{x1}+\widetilde b_{x1})\right)
-\bar u\partial _\xi \widetilde b_{x1}\nonumber\\
&&\quad\quad
-\frac{1}{2}\partial _\perp(\bar u_1 \widetilde b_1^*-\bar u_1^*\widetilde b_1)
+\bar b_1\partial_\xi\widetilde u_{x1}\nonumber\\
&&\quad\quad
\frac{1}{2}\partial_\perp(\bar b_1\widetilde b_1^*-\bar b_1^*\widetilde b_1)
+\frac{i}{R_i}\partial_{\xi\xi}\widetilde b_1 =0.
\label{o5/2:2b}
\end{eqnarray}
Similarly, the equation for $u_1$ leads to
\begin{eqnarray}
&&\partial _\tau \bar u_1+\partial_\perp\left(\beta(\bar\rho_1+j^2\bar\Phi)+\bar b_{x1}
+\langle\frac{|\widetilde b_1|^2}{2}\rangle\right)\nonumber\\
&&\quad +\frac{1}{2}
(\bar u_1\partial _\perp^* +\bar u_1^*\partial _\perp)\bar u_1
-\frac{\bar b_1}{2}(\partial_\perp^*\bar b_1-\partial_\perp\bar b_1^*)
=0,\nonumber \\
\label{o5/2:4a}
\end{eqnarray}
and 
\begin{eqnarray}
&&\partial_\tau \widetilde u_1+
\partial_\xi\left ((\widetilde\rho_1-\widetilde u_{x1}-\widetilde b_{x1})\widetilde b_1\right)+
(\bar\rho_1-\bar u_{x1}-\bar b_{x1})\partial_\xi \widetilde b_1\nonumber\\
&&\quad -\partial_\xi(\widetilde u_2 + \widetilde b_2)
-\frac{1}{2} (\bar u \partial_\perp ^* + \bar u^* \partial_\perp )\widetilde b 
-\frac{1}{2}(\widetilde b \partial_\perp ^* +\widetilde b^* \partial_\perp ) \bar u \nonumber\\
&&\quad +\partial_\perp\left(\beta(\widetilde\rho_1+j^2\widetilde\Phi)+\widetilde b_{x1}+\frac{|\widetilde b_1|^2
-\langle|\widetilde b_1|^2\rangle}{2}\right)\nonumber\\
&&\quad -\frac{\widetilde b_1}{2}(\partial_\perp^*\bar b_1-\partial_\perp\bar b_1^*)
-\frac{\bar b_1}{2}(\partial_\perp^*\widetilde b_1-\partial_\perp\widetilde b_1^*)
=0
\label{o5/2:4b}
\end{eqnarray}
where $\langle . \rangle$ denotes the mean value with respect to the
$\xi$ variable.
The solvability condition for 
eqs. (\ref{o5/2:2b}) and (\ref{o5/2:4b}) leads to
\begin{eqnarray}
&&\partial_\tau \widetilde b_1+
\partial_\xi\left((-\frac{\widetilde\rho_1}{2}+
\widetilde u_{x1}+\widetilde b_{x1})\widetilde b_1\right)
\nonumber\\
&&
\quad 
+(-\frac{\bar\rho_1}{2}+\bar u_{x1}+\bar b_{x1})\partial _\xi \widetilde b_1
+\frac{1}{2}\partial_\xi(\bar b_1\widetilde u_{x1})
\nonumber\\
&&
\quad
-\frac{1}{2}\partial_\perp\left(\beta(\widetilde\rho_1+j^2\widetilde\Phi)+\widetilde b_{x1}
+\frac{|\widetilde b_1|^2}{2}
-\langle\frac{|\widetilde b_1|^2}{2}\rangle\right)
\nonumber\\
&&
\quad 
+\frac{1}{2}(\bar u_1^*\partial_\perp
+\bar u_1\partial_\perp^*)\widetilde b_1
+\frac{1}{4}\bar b_1(\partial_\perp^*\widetilde b_1
-\partial_\perp\widetilde b_1^*)
\nonumber\\
&&
\quad
+\frac{1}{4}\widetilde b_1(\partial_\perp^*\bar b_1
-\partial_\perp\bar b_1^*)
-\frac{1}{4}\partial_\perp(\widetilde b_1^*\bar b_1-\widetilde b_1\bar b_1^*)
\nonumber\\
&&
\quad
+\frac{i}{2R_i}\partial_{\xi\xi}\widetilde b_1
=0,
\label{o5/2:5}
\end{eqnarray}
which generalizes the usual DNLS equation by the presence of
coupling to both longitudinal and transverse mean fields.

When pushing to the next order the asymptotic expansion for the longitudinal 
fields and combining the equations arising at successive orders $\epsilon^2$ 
and $\epsilon^3$ for these fields and $\epsilon^{3/2}$ and $\epsilon^{5/2}$
for the transverse mean fields, we obtain the self-gravitating version
of equations derived by \cite{Champ4}, which reads:
%
\begin{eqnarray}
&&\partial_\tau \widetilde b
+\partial_\xi\Big (\widetilde b(u_x+ b_x-\frac{ \rho}{2})
              +\frac{1}{2}\bar b(u_{x}+b_{x})\Big )\nonumber\\
&&\quad\quad- \frac{1}{2}\partial_\perp\Big(\beta(\widetilde\rho_1+j^2\widetilde\Phi) +
(1-\beta)\widetilde P\Big)\nonumber\\
&& \quad\quad +(\bar{\bf u}+\bar{\bf b})\cdot\mbox{\boldmath $\nabla$}\widetilde 
b
+\frac{i}{2R_i}\partial_{\xi\xi}\widetilde b
=0\label{sf:1b}\\
&&\partial _\tau \delta
+\frac{1}{\epsilon}\partial_\xi(u_x-\delta)
+\partial_\xi(\rho u_x)
+\frac{1}{2}\left(
\partial_\perp^*\left(\widetilde b( u_x-\delta)\right)+ 
{\rm c.c.}\right)\nonumber\\
&&\quad\quad
+\mbox{\boldmath $\nabla$}\cdot(\bar{\bf u}\delta+\bar{\bf b}u_{x})
- \frac{i}{2R_i}\partial_\xi
(\partial_\perp^*\widetilde b-\partial_\perp\widetilde b^*)
=0\label{sf:3b}\\
&&\partial_\tau  u_x-\frac{1}{\epsilon}\partial_\xi
\left(u_x-\beta(\rho_1+j^2\Phi)-\frac{|\widetilde b+\bar b|^2}{2}\right) \nonumber \\
&&\quad\quad +\partial_\xi\left (\frac{ u_x^2}{2}
+\left(\beta(\gamma-1)-1\right)\frac{\rho^2}{2}
-  b_x( b_x+u_x-\rho)\right )
\nonumber\\
&&\quad\quad
- \frac{1}{2}\left (\partial_\perp^*\left(\widetilde b(b_x+ u_x)\right)
+{\rm c.c.} \right )
\nonumber\\
&&\quad\quad +\mbox{\boldmath $\nabla$}\cdot(\bar{\bf u} u_x-\bar{\bf b}b_x)
=0\label{sf:5b}\\
&&\partial_\tau  b_x
+\frac{1}{\epsilon}\left (-\partial_\xi {b_x} 
+\mbox{\boldmath $\nabla$}\cdot{\bf u}\right )
\nonumber\\&&\quad\quad 
-\frac{1}{2}\left (\partial_\perp^*\widetilde b(u_x+b_x)+
{\rm c.c.}\right )
+\mbox{\boldmath $\nabla$}\cdot(\bar{\bf u}b_x-\bar{\bf b}u_x)\nonumber\\
&&\quad\quad+\frac{i}{2R_i}\partial_\xi
(\partial_\perp^*\widetilde b-\partial_\perp\widetilde b^*)
=0\label{sf:9b}\\
&&\partial_\tau \bar{\bf  u}-\frac{1}{\epsilon}\partial_\xi (\bar {\bf u}+\bar
{\bf b}) + \mbox{\boldmath $\nabla$}\left (\beta(\bar\rho_1+j^2\bar\Phi)
+\bar b_x
+\langle\frac{|\widetilde b|^2}{2}\rangle\right)\nonumber\\
&&\quad\quad + \bar {\bf u}\mbox{\boldmath $\cdot \nabla$}\bar {\bf u}
= ( \mbox{\boldmath $\nabla$}{\mbox{\boldmath $\times$}} \bar {\bf b} ) 
{\mbox{\boldmath $\times$}} \bar {\bf b} \label{sf:2}\\
&&\partial _\tau \bar {\bf b} -\frac{1}{\epsilon}\partial_\xi 
(\bar {\bf u}+\bar {\bf b})
-  \mbox{\boldmath $\nabla$}{\mbox{\boldmath $\times$}}
(\bar{\bf  u}{\mbox{\boldmath $\times$}} \bar{\bf b})  
=0\label{sf:10}\\
&&
\partial_{\xi} b_x +\frac{1}{2}\left(\partial_\perp(\bar b^*+\widetilde
b^*)+\partial_\perp^*(\bar b+\widetilde b)\right)=0\label{sf:11}\\
&&
\widetilde P= \frac{1}{2(1-\beta)} 
\left(2\widetilde b_x+|\widetilde b+\bar b|^2-\langle |\widetilde b+\bar b|^2
\rangle\right)\label{sf:12},
\end{eqnarray}
where $\delta=\rho -b_x$, $\bar u=\bar u_1+\epsilon\bar u_2$, $u_x= u_{x1}+\epsilon u_{x2}$, $b_x= b_{x1}+\epsilon b_{x2}$, $\delta= \delta_{1}+\epsilon\delta_{2}$, $\mbox{\boldmath $\nabla$}=(\partial _\eta,
\partial_\zeta)$, $\bar{\bf b}=(\bar b_y, \bar b_z)$ and 
$\bar{\bf u}=(\bar u_y, \bar u_z)$. Note that the equation for $\widetilde b$,
which results from a solvability condition, is kept unchanged. 
The above equations describe the nonlinear AW dynamics in the 
long-wavelength limit (Eq. (\ref{sf:1b})) and the two-dimensional hydrodynamics 
developing in planes perpendicular to the mean magnetic field 
(Eqs. (\ref{sf:2}) and (\ref{sf:10})). These equations also include
the non-linear dynamics of the magneto-sonic waves resulting
from the perturbation of the AW (Eqs. (\ref{sf:3b})-(\ref{sf:9b})).

Assuming that the mean quantities do not 
depend on the large longitudinal scale $\epsilon\xi$ and considering only 
the dominant contributions, except 
for the equation for $b_x$ where order $\epsilon$ contribution is kept
in order to allow magneto-sonic waves to develop as a consequence
of coupling between parallel propagating AW and transverse perturbations,  
we obtain    
\begin{eqnarray}
&&\widetilde u_x-\widetilde\delta=0\label{sf:3}\\
&&\widetilde
u_x-\beta(\widetilde\rho_1+j^2\widetilde\Phi)-\frac{|\widetilde b|^2-\langle|\widetilde b|^2\rangle
+\widetilde b\bar b^*+\widetilde b^*\bar b}{2}   =0 \label{sf:5}\\
&&\partial_\xi\widetilde b_x+\partial_\eta\widetilde b_y+\partial_\zeta\widetilde b_z=0
\label{sf:7}\\
&&\partial_\tau \bar\delta
+\left(\frac{1}{2}\partial_\perp^*\left[\langle\widetilde b(\widetilde u_x-\widetilde\delta)\rangle
+\bar u\bar\delta +\bar b\bar u_{x}\right] + c.c.\right)
=0\label{sf:4}\\
&&\partial_\tau  \bar u_x 
+\left(\frac{1}{2}\partial_\perp^*\left[\bar u\bar u_x
-\langle\widetilde b(\widetilde b_x+\widetilde u_x)\rangle-
\bar b\bar b_x\right] + c.c.\right)\nonumber\\
&&\quad\quad=0\label{sf:6}\\
&&\partial_\tau \bar b_x
+\frac{1}{2\epsilon}(\partial_\perp\bar u^*+\partial_\perp^*\bar u)\nonumber\\
&&\quad\quad -\frac{1}{2}\left[
\langle\partial_\perp^*(\widetilde b(\widetilde u_x+\widetilde b_x))\rangle\right]\nonumber\\
&&\quad\quad -\frac{1}{2}\left[\partial_\perp^*(\bar b\bar u_x)
-\partial_\perp^*(\bar u\bar b_x)+c.c.\right]
=0\label{sf:9}\\
&&\partial_\tau \bar u
+ \partial_\perp\left(\beta(\bar\rho_1+j^2\bar\Phi)+\bar b_x
+\langle\frac{|\widetilde b|^2}{2}\rangle\right)\nonumber\\
&&+\frac{1}{2}(\bar u\partial_\perp^*\bar u+\bar u^*\partial_\perp\bar u)
-\frac{1}{2}(\bar b\partial_\perp^*\bar b-\bar b\partial_\perp\bar b^*)
=0
\label{sf:21}\\
&&\partial _\tau \bar b 
+\frac{1}{2}\partial_\perp(\bar u\bar b^*-\bar u^*\bar b)=0\label{sf:101}\\
&&\partial_\eta\bar b_y+\partial_\zeta\bar
b_z=0,\label{sf:8} 
\end{eqnarray} 
where oscillating and mean contributions have been separated.
Previous equations can be used to rewrite the solvability condition 
as 
\begin{eqnarray}
&&
 \partial_\tau \widetilde b +\partial_\xi \left (
       \frac{1}{2}(\widetilde b+\bar b)\widetilde P
        +(\bar u_x+\frac{1}{2}\bar b_x-\frac{1}{2}\bar\delta)\widetilde b\right )
   -\frac{1}{2}\partial_\perp \widetilde P\nonumber\\
&&\qquad +(\bar{\bf u}+ \bar{\bf b})\cdot\mbox{\boldmath $\nabla$}\widetilde b
       +\frac{i}{2R_i} \partial_{\xi\xi}\widetilde b=0\label{proc:45}
\end{eqnarray}

\end{document}